\documentstyle[epsfig]{mn2e}

\def\gap{\;\rlap{\lower 2.5pt
 \hbox{$\sim$}}\raise 1.5pt\hbox{$>$}\;}
\def\lap{\;\rlap{\lower 2.5pt
   \hbox{$\sim$}}\raise 1.5pt\hbox{$<$}\;}
\def\gsim{\;\rlap{\lower 2.5pt
 \hbox{$\sim$}}\raise 1.5pt\hbox{$>$}\;}
\def\lsim{\;\rlap{\lower 2.5pt
 \hbox{$\sim$}}\raise 1.5pt\hbox{$<$}\;}
\def\msun{{\rm\,M_\odot}}

\def\cm{{\rm\,cm}}
\def\sec{{\rm\,s}}

\def\etal{{\rm\,et \ al. }}
\def\sr{{\rm\,sr}}

\def\cm{{\rm\,cm}}

\def\kpc{{\rm\,kpc}}
\def\pc{{\rm\,pc}}

\def\m{{\rm\,m}}
\def\GeV{{\rm\,GeV}}

\def\sec{{\rm\,s}}
\def\sr{{\rm\,sr}}

\def\Ba{B$_{z_0}$} 
\def\Bb{B$_{z_0,5\sigma}$}
\def\Bc{B$_{z_c}$}
\def\Bd{B$_{z_c,5\sigma}$}
\def\Ea{ENS$_{z_0}$}
\def\Eb{ENS$_{z_c}$}
\def\Bref{B$_{ref,z_0}$} 
\def\Bzref{B$_{ref,z_c}$}
\def\spose#1{\hbox to 0pt{#1\hss}}
\def\lta{\mathrel{\spose{\lower 3pt\hbox{$\mathchar''218$}}
     \raise 2.0pt\hbox{$\mathchar''13C$}}}
\def\gta{\mathrel{\spose{\lower 3pt\hbox{$\mathchar''218$}}
     \raise 2.0pt\hbox{$\mathchar''13E$}}}
\newcommand{\beq}{\begin{equation}}
\newcommand{\eeq}{\end{equation}}
\newcommand{\be}{\begin{equation}}
\newcommand{\ee}{\end{equation}}

\newcommand{\ls}{\mathrel{\raise1.16pt\hbox{$<$}\kern-7.0pt 
\lower3.06pt\hbox{{$\scriptstyle \sim$}}}}         
\newcommand{\gs}{\mathrel{\raise1.16pt\hbox{$>$}\kern-7.0pt 
\lower3.06pt\hbox{{$\scriptstyle \sim$}}}}         

\long\def\comment#1{}

\def\msun{M_{\odot}}
\def\fun#1#2{\lower3.6pt\vbox{\baselineskip0pt\lineskip.9pt
  \ialign{$\mathsurround=0pt#1\hfil##\hfil$\crcr#2\crcr\sim\crcr}}}
\def\lap{\mathrel{\mathpalette\fun <}}
\def\gap{\mathrel{\mathpalette\fun >}}
\newcommand{\ba}{\begin{eqnarray}}
\newcommand{\ea}{\end{eqnarray}}

\def\rsun{{R_\odot}}    
\title{Dark Matter Annihilation in Substructures Revised}
\author[L. Pieri, G. Bertone \& E. Branchini]{Lidia Pieri$^{1,2}$,
Gianfranco Bertone$^3$ and
Enzo Branchini$^4$ \\
$^1$Istituto Nazionale di Astrofisica - Osservatorio di Padova, \\Vicolo dell'Osservatorio 5, 35122 Padova, Italy \\
$^2$Istituto Nazionale di Fisica Nucleare - Sezione di Padova, \\ Via Marzolo 8, 35131 Padova, Italy \\
$^3$Institut d'Astrophysique de Paris, UMR 7095-CNRS,\\Universit\'e Pierre et Marie Curie, 98bis boulevard Arago, 75014 Paris, France \\
$^4$Department of Physics, Universit\`a di Roma Tre,\\ Via della Vasca Navale 84, 00146, Rome, Italy}
\begin{document}

\maketitle
\begin{abstract}
Upcoming $\gamma$-ray satellites will search for Dark Matter annihilations in Milky Way substructures (or 'clumps'). The prospects for detecting these objects strongly depend on the assumptions made on the distribution of Dark Matter in substructures, and on the distribution of substructures in the Milky Way halo. By adopting simplified, yet rather extreme, prescriptions for these quantities, we compute the number of sources that can be detected with upcoming experiments such as GLAST, and show that, for the most optimistic particle physics setup ($m_\chi=40$ GeV and annihilation cross section $\sigma v = 3 \times 10^{-26}$ cm$^3$ s$^{-1}$), the result ranges from zero to $\sim$ hundred sources, all with mass above  $10^{5}M\odot$.
However, for a fiducial DM candidate with mass $m_\chi=100$ GeV and $\sigma v = 10^{-26}$ cm$^3$ s$^{-1}$, at most a handful of large mass substructures can be detected at $5 \sigma$, with a 1-year exposure time, by a GLAST-like experiment.     
Scenarios where micro-clumps (i.e. clumps with mass as small as $10^{-6}M\odot$) can be detected are severely constrained by the diffuse $\gamma$-ray background detected by EGRET. 

\end{abstract}

\noindent

\section{Introduction}

Indirect Dark Matter [DM] searches are based on the detection 
of secondary particles and radiation produced by the 
self-annihilation of DM particles (Bergstr\"om 2000, Bertone \etal 2005a). 

Although the predicted annihilation flux is typically affected by large 
astrophysical uncertainties, the detection of multiwavelenght photons,
neutrinos or anti-matter from regions with high DM density would be of
paramount importance for the identification of DM particles. In fact, 
accelerator searches of Physics beyond the Standard Model at the Large 
Hadron Collider, will not necessarily unveil the nature 
of DM, even if new particles are discovered, due to the difficulties
associated with the reconstruction of the cosmological abundance of the 
newly discovered particles (e.g. Baltz \etal 2006a, Nojiri \etal 2005). 
At the same time, DM particles could 
have small enough couplings to nucleons, to lead to null searches in 
direct detection experiments (see e.g. Mu${\rm \tilde n}$oz 2003 and references therein).

In the framework of indirect DM searches, several strategies have 
thus been devised, in order to obtain {\it conclusive} evidence from astrophysical 
observations. For instance, one could search for peculiar features, such
as lines or sharp cut-offs, in the $\gamma$-ray spectrum. Although for 
commonly studied DM candidates there are no tree level processes for 
direct annihilation into photons, loop-level processes to $\gamma \gamma$ and 
$\gamma Z^0$ may produce detectable lines at an energy equal to the DM particle mass 
(see e.g. Bergstr\"om \& Ullio 1997, Ullio \& Bergstr\"om 1998, Gounaris \etal 2003, Bergstr\"om \etal 2005a).
Other spectral features 
may help distinguishing the DM annihilation signal from ordinary astrophysical 
sources (Bergstr\"om \etal 2005b, Bergstr\"om \etal 2005c; see also the discussion in Baltz \etal 2006b).
Alternatively, one can search for annihilation 
radiation from regions characterized by large concentrations 
of DM, but very few baryons, such as DM substructures in the Milky Way [MW hereafter] halo,
including dwarf galaxies (Baltz \etal 2000, Tasitsiomi \etal 2003, Pieri \& Branchini 2003, Evans \etal 2003, Tyler 2002, 
Colafrancesco \etal 2007, Bergstr\"om \& Hooper 2006) and DM mini-spikes around Intermediate Mass Black 
Holes (Bertone 2006, Bertone \etal 2005b, Horiuchi \& Ando 2006, Fornasa \etal 2007, Brun \etal 2007).
Finally, DM annihilation features  can be detected in the energy spectrum and angular distribution
of the cosmic $\gamma$-ray background (Bergstr\"om, Edsj\"o \& Ullio 2001, Ando \& Komatsu 2006).

In the popular Cold DM scenario, gravitational instabilities 
lead to the formation of a wealth of virialized structures, the DM haloes,  
spanning a huge range of masses, from the largest clusters of galaxies  of $\sim 10^{15} M_\odot$ 
down to Earth-size clumps of $\sim 10^{-6} M_\odot$ (Green \etal 2004, Green \etal 2005).
Although the detectability
 of individual DM substructures, or "clumps"
has been widely discussed in literature, the number of detectable clumps 
with a GLAST-like experiment, at $5\sigma$ in 1 year and for a WIMP DM 
particle is highly uncertain, ranging from $\lsim 1$ (Koushiappas \etal 2004) 
to more than 50 (Baltz 2006b) for large mass haloes,
while for microhaloes (i.e. clumps with a mass as small as 
$10^{-6} M_\odot$) the predictions range from no detectable objects
(Pieri \etal 2005)
to a large number of detectable objects, with a fraction of them 
exhibiting a large proper motion (Koushiappas 2006).
The apparent inconsistency of the results published so far, is actually due to 
the different assumptions 
that different groups adopt for the physical quantities that 
regulate the number and the annihilation
"brightness" of DM clumps.
In particular, even in the  context  of the benchmark density
 profile introduced by Navarro, Frenk and White 1996 [NFW],
the results crucially depend on the substructures mass function, 
their distribution within the halo host and their 
virial concentration $c(M,z)$ which 
is a function of mass and of collapse redshift of DM clumps.

The paper is organized as follows: in Sec.~\ref{sec:models}, we describe 
the model we have adopted for the smooth component of the Galactic halo,
and introduce the eight different models for DM substructures that will be discussed
in the rest of the paper. In Sec.\ref{sec:flux} we estimate the contribution
to the $\gamma$-ray flux due to the smooth Galactic halo, unresolved DM clumps,
and resolved (detectable) clumps. In Sec.\ref{sec:detection}, we study the
prospects for detection of substructures with upcoming experiments such 
as GLAST, and in Sec.\ref{sec:conclusions} we discuss the results and present 
our conclusions.

\section{Modeling the Galactic halo and its substructures} 
\label{sec:models}
High resolution N-body experiments indicate that a large fraction of
the mass within Dark Matter haloes is in the form of virialized
subhaloes in all resolved mass scales.
Their annihilation signal, which adds to that of the smooth Galactic component 
could be significant
(Stoher \etal 2003, Diemand \etal 2006, Diemand \etal 2007a).
A precise modeling of both the smooth DM distribution (the diffuse galactic component)
and the subhalo population within (the clumpy component) is therefore mandatory
to assess the possibility of detecting DM subhaloes through their annihilation signal.
High resolution numerical simulations enable to study gravitationally bound subhaloes 
with $M_{\rm SH} \ge 10^{-6} \ M_{\rm Halo}$, where $\ M_{\rm Halo}$ is the mass of the host,
 and therefore cannot resolve 
substructures  in a MW-size halo all the way down to 
$\sim 10^{-6} \msun$. 
In fact, the smallest substructures of interest must be studied within host haloes
with mass $\sim 0.1 \msun$  and only at very large redshift (Diemand \etal 2006).
At $z=0$ and within a  MW-host the smallest subhaloes that we resolve have 
masses $\ge 10^6 \msun$ (Diemand \etal 2007a)
As a consequence the spatial distribution, mass function and internal structure
of Galactic subhaloes can only be interpolated from the results of 
several numerical experiments spanning a large range of masses and redshifts
using self-similarity  arguments.
This interpolation is affected by a number of uncertainties that 
we account for by exploring different models that meet the numerical constraints. 

\subsection{The diffuse Galactic component} 
\label{sec:diffgal} 
The recent ``Via Lactea'' high resolution simulation
(Diemand \etal 2007a)  shows that the density profile 
of a MW-sized DM halo is consistent to within 10\% 
with the NFW profile that we adopt here:
\begin{eqnarray}
\rho_\chi(r) & = & \frac{\rho_s}{\left( \frac{r}{r_s} 
\right) \left(1 + \frac{r}{r_s} \right)^2 }, 
\ \ \ r > r_{min} \nonumber  \\ \nonumber
\rho_\chi(r) & = & \rho_\chi(r_{min}), \ \ \ r \leq r_{min} \\
\end{eqnarray}
where $r$ is the distance from the halo center. This profile depends on two free parameters,
the scale density,  $\rho_s$, and the scale radius, $r_s$, that are related to each other 
by the virial mass of the halo, $M_{h}$. The latter is the mass enclosed in a sphere
with radius $r_{vir}$  within which the mean density  is $200$ times above critical.
A different definition of $r_{vir}$ would not change the DM profile nor our results. .
Finally, we adopt a small core radius $r_{min}=10^{-8} \kpc$. 

An important shape parameter that characterize the density profile is the virial
concentration defined as the ratio between the virial radius and the scale radius,
$c \equiv r_{vir} (M_{h}) /r_s $. 
Theoretical considerations corroborated by numerical experiments show that
a relation exists between the mass of a halo, its collapse redshift, $z_{\rm coll}$, and 
concentration parameter $c$.  The collapse redshift is defined as in  Bullock \etal 2001,
as the epoch in which a mass scale $M_{h}$ breaks into the nonlinear regime, i.e. 
when $\sigma(M_{h})D(z_{\rm coll}) \sim 1$, where 
$\sigma(M_{h})$ is the present linear theory amplitude of mass fluctuation on the scale 
$M_{h}$ and $D(z_{\rm coll})$ is the linear theory growth factor at the redshift $z_{\rm coll}$.
The two models proposed by Bullock \etal 2001 [B01] and Eke \etal 2001 [ENS01] are consistent for
masses larger than $\sim 10^{9} \msun$ and in this paper we use the concentration parameter 
by B01 to model the diffuse Galactic component. On the contrary, for smaller masses 
ENS and B01 predictions become very different  and we will have to consider both of them to 
model the subhaloes annihilation signal.

Since a sizable fraction of the mass in the MW is in form of virialized subhaloes, there is not
a unique way to determine $\rho_s$, and $r_s$ of the host MW halo.
For this reason we adopted two different procedures.
In the first one we have computed $\rho_s$, and $r_s$  as if the total mass of the system,
including the clumpy component, were in fact smoothly distributed in the Galactic halo. 
In the second we have considered the total mass of the system 
 to determine the concentration parameter but have 
used the diffuse component alone to relate $\rho_s$ to $r_s$ .
Having checked that both procedures give similar predictions for the 
probability of subhalo detection  (Section \ref{sec:flux}), in the following we will only discuss models based on the second approach. For the Milky Way we have used a virial mass $M_h = 10^{12} \msun$ and a
concentration parameter $c_{vir} \sim 9.8$.

\subsection{The clumpy component} 

\begin{table}
\begin{center}
\begin{tabular}{|c|c|c|c|c|c|}
\hline 
  & \multicolumn{4}{|c|}{Subhalo Models} \\ \hline
Model &  \multicolumn{2}{|c|}{$c(M)$}  & $z_{\rm coll}$ & $c$  \\ 
name & $> 10^{4} M_\odot$ &  $10^{-6}M_\odot$  & $10^{-6}M_\odot$ & $10^{-6}M_\odot$ \\ \hline 
\Bref & B01 & B01 & 0 & 63 \\ \hline
\Ba  & B01 & DMS05 & $0$ & 80 \\ \hline
\Bb  & B01 & DMS05-$5\sigma$ & $0$ & 400 \\ \hline
\Bzref & B01 & B01 & B01 & $\sim 3.7$ \\ \hline
\Bc & B01 & DMS05 & DMS05 & $\sim 1$ \\ \hline
\Bd & B01 & DMS05-$5\sigma$& DMS05 & $\sim 0.6$ \\ \hline
\Ea & ENS01 & DMS05 & 0 & 80 \\ \hline
\Eb & ENS01 & DMS05 & DMS05 & $\sim 1$\\ \hline
\hline 
\end{tabular}
\label{tab1}
\caption{Halo model parameters. Column 1: Model name. Column 2: Halo density profile for $M > 10^{4} \msun$.
Column 3. Halo density profile benchmark for $M=10^{-6} \msun$,  used as normalization. Column 4: Collapse redshift model, used as normalization. Column 5: 
Concentration parameter at $10^{-6} \msun$. For all the models
we have adopted an NFW profile.
}
\end{center}
\end{table}

To account  for the presence of a 
population of DM subhaloes and investigate their 
effect on the annihilation signal,
we need to specify their mass
spectrum, spatial distribution and 
density profile.

\subsubsection{The Mass function and the Spatial Distribution of Clumps} 

High resolution N-body experiments show that the mass function of both isolated field haloes and 
subhaloes is well approximated by a power law
\begin{equation}  
{\rm d} n(M)/{\rm d ln}(M) \propto M^{-\alpha},
\label{mass}  
\end{equation}  
with $\alpha$ = 1,
 independently of the host halo mass, over the large redshift range
$z=[0,75]$, and mass intervals $M=[10^{-6},10^{10}]\msun$ 
(Jenkins \etal 2001, 
Moore \etal 2001
Diemand \etal 2004,
Gao \etal 2005,
Reed \etal 2005,
Diemand \etal 2006).
Self-similarity is preserved  at the present epoch
down to the smallest masses if subhaloes  
survive gravitational disturbances during early merger processes and 
late tidal disruption from stellar encounters. 

Analytical arguments have been given against (Zhao et al. 2005)
or in support of this hypothesis (Moore et al. 2005, 
Brezinsky et al. 2006). Moreover,  currently resolved mass functions 
in numerical experiments suffer from  dynamical friction at the high mass end
which could steepen the halo mass function.
Since changing the halo mass function slope might have a non-negligible impact
on our analysis,  we adopt a power-law index $\alpha=1$ as a reference case 
but also explore two shallower subhalo mass functions with  $\alpha = 0.9$ and $\alpha = 0.95$.
All plots in this paper refer to the reference case $\alpha=1$ and discuss the effect 
of adopting shallower slopes in the text.

As far as the spatial distribution of subhalos inside our Galaxy is concerned,
we follow the indications of the numerical experiment of  
Reed \etal 2005
and assume that the subhalo distribution 
traces that of the underlying host mass
from $r_{vir}$ and down to a minimum radius, $r_{min}(M)$,
within which subhaloes are efficiently destroyed by gravitational 
interactions.
We explicitly assume spherical symmetry and 
we ignore the possibility, indicated by some numerical 
experiments, that the radial distribution of subhalos might be 
more extended than that of the dark matter.
  
Folding  these indications together we model 
the number density of 
subhaloes per unit mass at a distance $R$ from the GC as:   
\begin{equation}  
\rho_{sh}(M,R) = A M^{-2} \frac{\theta (R - r_{min}(M))}{(R/r_s^{MW}) 
(1 + R/r_s^{MW})^{2}} \msun^{-1}  \kpc^{-3},  
\label{rho}  
\end{equation}  
where $r_s^{MW}$ is the scale radius of our Galaxy and 
the effect of tidal disruption is accounted for by the
Heaviside step function $\theta (r - r_{min}(M))$.
To determine the tidal radius, $r_{min}(M)$, we follow the Roche 
criterion and compute it  as the minimum distance 
at which the subhalo self gravity at $r_s$ equals the gravity
pull of the halo host computed at the orbital radius
of the subhalo. 
As a result $r_{min}(M)$ is an increasing function of  
the subhalo mass, implying that no subhaloes survive within  
$r_{min}(10^{-6} \msun) \sim 200 \pc$.  

To normalize eq.~\ref{rho}, we again refer to numerical simulations
that show that [5-10]\% of the MW mass 
is distributed in subhaloes with masses in the range  
$10^{7}-10^{10} \msun$ (Diemand \etal 2005,
Diemand \etal 2007)
In the following we use the optimistic value of 10 \% and note
that assuming 5\%, instead, would decrease the probability 
of subhalo detection by a factor 2.
With this normalization about 53\% of 
the MW mass is condensed within
$\sim 1.5 \times 10^{16}$ subhaloes with masses in the range 
[$10^{-6},10^{10}] \msun$,
whose abundance in the solar neighborhood is remarkably high
($\sim 100 \pc^{-3}$).
The remaining 47\% constitutes the diffuse galactic component that
is assumed to follow a smooth NFW profile. We do not account here 
for the  presence of mass debris streams resulting from tidal stripping
since these structures, characterized by a mild density contrast, would not
contribute significantly to the annihilation flux.

\subsubsection{Density Profile.} Finally, we need to specify the density profile for the substructures.
Constraints from numerical models only applies to limited mass ranges
at very different epochs. 
At $z=0$ Diemand \etal 2007b find that the velocity profile of 
Galactic subhaloes above $4 \times 10^6 \msun$ in the ``Via Lactea'' simulation are
well fitted by the NFW model.
This result is also valid for the much smaller substructures with masses in the range
$[10^{-6},4\times10^{-3}]$ $\msun$ that populate a parent halo of $0.014 \msun$
at $z=86$  (Diemand \etal 2006).
A large fraction of these small substructures do not 
survive the early stage of hierarchical merging and
late tidal interaction with stellar encounters (Zhao \etal 2005).
The $\sim 10^{16}$ survivors 
suffer from significant mass loss.  Presumably this
modifies their original NFW density profile that, however,  
seems to be preserved in the innermost region where most of the
annihilation signal originates from (Kazantzidis \etal 2004). 
 
Yet these constraints from numerical experiments do not uniquely
define the subhalo density profiles.
Therefore, instead of relying on 
a single model profile we will explore several of them  
in an attempt of bracketing the theoretical uncertainties.

\begin{figure}
\epsfig{file=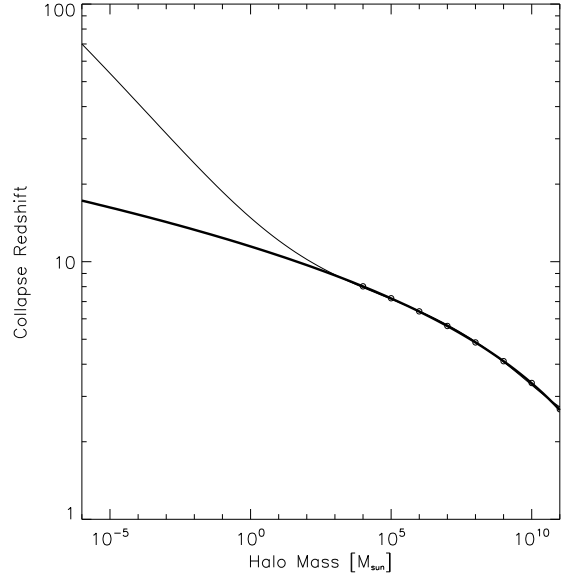,width=0.65\textwidth}
\caption{Collapse redshift, $z_c$, as a function of the halo mass for model B01 (points) for $M>10^{4} \msun$. The thin line shows the extrapolation at low masses normalized to the DMS05 value for $z_c(10^{-6} \msun)$. The thick line shown a na\"ive extrapolation of the B01 model at low masses.
}
\label{fig1}
\end{figure}

\begin{figure}
\epsfig{file=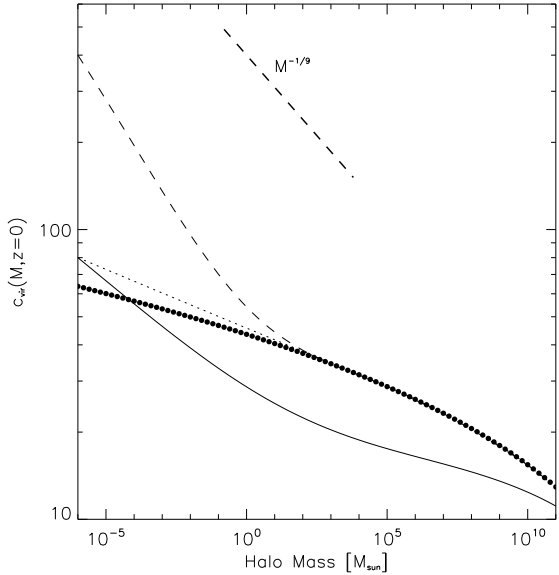,width=0.65\textwidth}
\caption{Concentration parameters as a function of halo mass
 at $z=0$ computed for the \Ea (solid),  \Ba (dotted) and the 
\Bb (dashed) model described in the text. Filled circles show the B01 model na\"ively extrapolated at low masses (\Bref model). 
}
\label{fig2}
\end{figure}

All models that we have considered, and that are listed in 
Table 1, assume that subhaloes have the same NFW density profile
as their massive host but with different concentration parameters.
These models have been flagged with the following prescriptions:
\begin{itemize}
\item {\it B}-models assume the $c(M)$ relation of B01 for $M > 10^4 \msun$;
\item {\it ENS}-models, use the ENS01 model for $M > 10^4 \msun$;
\item In models flagged as $z_0$ the low mass extrapolation of the concentration
parameter is normalized to that of field haloes of $10^{-6} \msun$ measured 
in the numerical simulation of Diemand \etal 2005 (DMS05) at $z=26$ and
linearly extrapolated at $z=0$, i.e.
$c(10^{-6} \msun, z=0)=c(10^{-6} \msun, z=26) \times (1+26)$. 
The underlying assumption is that, as in the Press-Schechter approach,
all existing haloes have just formed, and the $(1+26)$
scaling is required to account for the change in the mean density  
between $z=26$ and $z=0$;
\item Models flagged as $z_c$ assume instead that, once formed, 
subhaloes that survive to $z=0$ do not change their density profile.
Therefore for each subhalo of mass $M$ we determine its collapse redshift $z_c$
(defined as in B01 for $M > 10^4 \msun$) and compute its concentration parameter accordingly, as
$c(M, z=z_c)=c(M, z=0) / (1+z_c)$, where
$c(M,z=0)$ is computed as in the $z0$ case.
The small mass normalization is the same as in models $z0$, rescaled for the smallest masses
collapse redshift $z_c(M=10^{-6} \msun) = 70$ 
suggested by DMS05 and used by Koushiappas 2006.
In Fig.~\ref{fig1} we show the collapse redshift as a function of the mass
adopted in the $ z_c$-models (thin line);
\item Models flagged $5 \sigma$ assume that all existing subhaloes 
with mass $M=10^{-6} \msun$  form at the $5\sigma$ peaks of
the density field, which has the effect of increasing the normalization to
$c(10^{-6} \msun, z=0)=400$;
\item Models flagged as $ref,z0$ and $ref,zc$ are identical to the corresponding $z0$ and $zc$ models for  $M > 10^{4} \msun$. Indeed, they use a na\"ive extrapolation of the B01 model at low masses, both for the collapse redshift (thick line in Fig. \ref{fig1}) and for the concentration parameter (filled circles in Fig. \ref{fig2}). Though this extrapolation is not supported by numerical simulations, it intuitively reflects the theoretcal flattering of the $\sigma(M)$ curve at low masses.
\end{itemize}
The $c(M)$ profile for each of the $z0$ models is shown in Fig.\ref{fig2}. \\

Finally, as pointed out in B01, the $c(M)$ relation is not deterministic. Instead,
for a fixed mass, the probability of a given value for $c(M)$ is well described by a lognormal distribution
\begin{equation}
P(c(M)) = \frac{1}{\sqrt{2 \pi} \sigma_c \bar c(M)} \, e^{- \left ( \frac{\ln(c(M))-\ln(\bar c(M))}{4 \sigma_c} \right )^2},
\label{pcvir}
\end{equation}
where the mean $\bar c= c(M)$ is the concentration parameter of B01 or ENS and 
the dispersion $\sigma_c$ = 0.24 does not depend  on the halo mass (B01).
In the following, we include  this  lognormal scatter in all 
models described in Table 1.

We will not consider in this paper the possibility that subhaloes might contain sub-substructres
(Strigari \etal 2007, Diemand 2007a) and that their concentration (and scatter) might 
depend on the distance from the Galactic Centre (Diemand \etal 2007b).
Indeed, such features are found in numerical simulations capable of resolving 
large sub-haloes ($M>10^{6} \msun$) but there is no evidence whether they also 
apply  to much smaller haloes, with masses down to $10^{-6} \msun$.

\section{$\gamma$-ray flux from annihilation in DM clumps}
\label{sec:flux}
The photon flux from neutralino annihilation in galactic
subhaloes, from a direction in the sky making an angle
$\psi$ from the Galactic Center (GC), and observed by a
detector with angular resolution $\theta$, can be factorized into 
a term depending only on particle physics parameters,
$d \Phi^{\rm PP}/dE_\gamma$ and a term depending only
on cosmological quantities, $\Phi^{\rm cosmo}(\psi, \theta)$:
\begin{equation}
\frac{d \Phi_\gamma}{dE_\gamma}(E_\gamma, \psi, \theta) =
\frac{d \Phi^{\rm PP}} {dE_\gamma}(E_\gamma) \times \Phi^{\rm
cosmo}(\psi, \theta)
\label{flussodef}
\end{equation}

\subsection{Particle physics contribution}
\label{fluxpp}
\begin{figure}
\epsfig{file=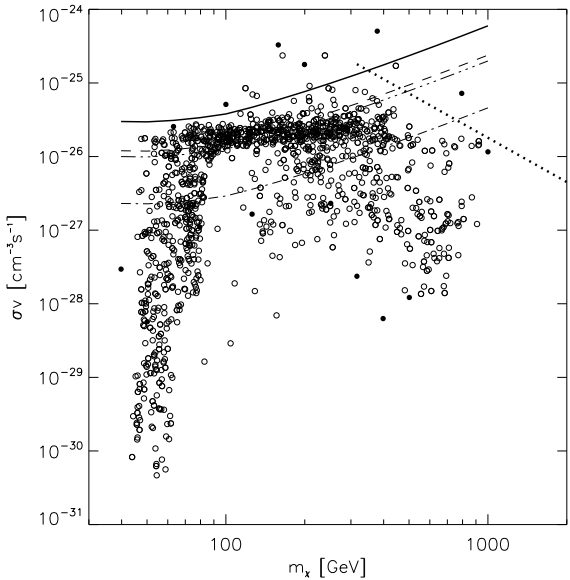,width=0.65\textwidth}
\caption{
Exclusion plot in the $(\sigma_{\rm ann} v, m_\chi)$ plane. The solid line corresponds 
to our best case particle physics scenario, adopted for all the {\it z0} models. 
Models above the dashed (long dot-dashed, dot-dashed) curve violate
the EGRET extra galactic background (EGB) flux constraints in scenario \Eb (\Bc,\Bd) for a mass function slope $\alpha=1$. 
For comparison, we show with filled (empty) dots a scan of SUSY models with 
relic density within 2 (5) standard deviations from the WMAP+SDSS value. 
The dotted line corresponds to the UED model. Supersymmetric models were obtained using the DARKSUSY package (Gondolo \etal, 2004)
}
\label{sigmavmchi}
\end{figure}
The first factor of Eq.\ref{flussodef} can be written as: 
\begin{equation}
\frac{d \Phi^{\rm PP}}{dE_\gamma}(E_\gamma) =  
  \frac{1}{4 \pi} \frac{\sigma_{\rm ann} v }{2 m^2_\chi} \cdot 
\sum_{f} \frac{d N^f_\gamma}{d E_\gamma} B_f  
\label{flussosusy}
\end{equation}
where $m_\chi$ denotes the Dark Matter particle mass
and $d N^f_\gamma / dE_\gamma$ is the differential photon
spectrum per annihilation relative to the 
final state $f$, with branching ratio $B_f$. Although the nature of the 
DM particle is unknown, we can make an educated guess on the
physical parameters entering in the above equation. The most commonly 
discussed DM candidates are the so-called
neutralinos, arising in Super-symmetric extensions of the Standard Model
of particle physics [SUSY], and the $B^{(1)}$ particles, first excitation of the
hypercharge gauge boson in theories with Universal Extra Dimensions [UED]
(see Bergstr\"om 2000, Bertone \etal 2005b and references therein).
Typical values for the mass of these candidates range from 
$\sim 50$ GeV up to several TeV. The annihilation cross section
can be as high as $ \sigma_{\rm ann} v  = 3 \times 10^{-26}$cm$^3$s$^{-1}$, as
appropriate for thermal relics that satisfy the cosmological constraints 
on the present abundance of Dark Matter in the Universe. However, we note 
that the annihilation 
cross section can be much smaller, as the appropriate relic density 
can be achieved through processes such as 
co-annihilations (Bergstr\"om 2000, Bertone \etal 2005a). 
This can be seen from Fig.~\ref{sigmavmchi} that shows the range 
of  $\sigma_{\rm ann} v $ and  $m_\chi$  allowed in the 
UED and SUSY models: solid (empty) circles correspond to models with relic density within 2 (5) 
standard deviations from the WMAP+SDSS suggested value 
$\Omega_{DM} h^2 = 0.1050^{+0.0041}_{-0.0040}$, where as usual $\Omega_{DM}$
is the DM density in units of the critical density, and $h$ is the Hubble
parameter in units of 100 km s$^{-1}$ kpc$^{-1}$, 
$h=0.730^{+0.019}_{-0.019}$ (Tegmark \etal 2006).

In order to optimize the prospects for detection, we  
adopt here a very low value for the particle mass, $m_\chi=40$ GeV, together with a 
high annihilation cross section $ \sigma_{\rm ann} v  = 3 \times 10^{-26}$cm$^3$s$^{-1}$.
As for the nature of the DM particle, we assume here a $100\%$ 
branching ratio in $b \bar b$ and the $d N^{b \bar b}_\gamma / dE_\gamma$ 
functional form of Fornengo \etal 2004. The results
can be rescaled for any other candidate, although in most cases the
photon spectrum arising from annihilations yields similar results.

\subsection{Annihilation flux from diffuse matter and unresolved clumps}
\begin{figure}
\epsfig{file=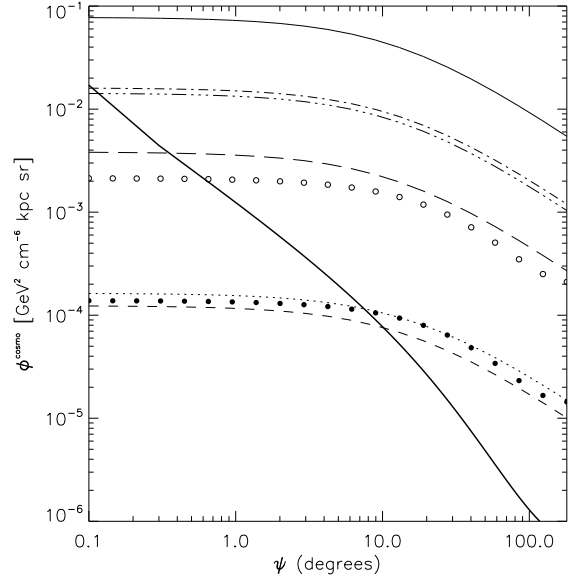,width=0.65\textwidth}
\caption{ $\Phi^{\rm cosmo}$ as a function of the angular distance from the Galactic center, for the MW smooth component (solid thick line) and for unresolved clumps in model \Bd (solid thin line), \Bc (dot-dashed), \Eb (long dot-dashed), \Bb (long dashed), \Bzref (open circles), \Ba (dotted), \Bref (filled circles) and \Ea (dashed). The mass function slope is $\alpha=1$.}
\label{fig3}
\end{figure}

The contribution of unresolved substructures to the annihilation signal is
given by
$$ 
\Phi^{\rm cosmo}(\psi, \Delta \Omega) = \int_M d M \int_c d c \int \int_{\Delta \Omega}
d \theta d \phi \int_{\rm l.o.s}  d\lambda
$$
$$
[ \rho_{sh}(M,R(\rsun, \lambda, \psi, \theta, \phi)) \times P(c) \times
$$
\begin{equation}
\times \Phi^{\rm cosmo}_{halo}(M,c,r(\lambda, \lambda ', \psi,\theta ', \phi ')) \times J(x,y,z|\lambda,\theta, \phi) ]
\label{smoothphicosmo}
\end{equation}
where $\Delta \Omega$ is the solid angle
of observation pointing in the direction of observation $\psi$ and defined by the
angular resolution of the detector $\theta$; $J(x,y,z|\lambda,\theta, \phi)$ 
is the Jacobian determinant; $R$ is the galactocentric distance, which, inside the cone, can be written as a 
function of the line of sight ($\lambda$) and the solid angle ($\theta$ and $\phi$) 
coordinates and the pointing angle $\psi$ through the relation  $R = \sqrt{\lambda^2 + \rsun^2 -2
\lambda \rsun C}$, where $\rsun$ is the distance of the Sun from the
Galactic Center and $C=\cos(\theta) \cos(\psi)-\cos(\phi) \sin(\theta) \sin(\psi)$; $r$ is the radial coordinate inside the single subhalo located at distance $\lambda$ from the observer along the line of sight defined by $\psi$ and contributing to the diffuse emission. \\
The expression
$$
\Phi^{\rm cosmo}_{halo}(M,c,r) = \int \int_{\Delta \Omega}  
d \phi ' d \theta '  \int_{\rm l.o.s} d\lambda '
$$
\begin{equation}
\left [ \frac{\rho_{\chi}^2 (M,c,r(\lambda, \lambda ',\psi,\theta ' \phi '))} {\lambda^{2}} J(x,y,z|\lambda ',\theta ' \phi ') \right] \, ;
\label{singlehalophicosmo}
\end{equation}
describes the emission from such a subhalo. 
Here, $\rho_\chi(M,c,r)$ is the Dark Matter density profile inside the halo.

By numerically integrating Eq. \ref{smoothphicosmo}, we estimate the 
contribution to  $\Phi^{\rm cosmo}$ from unresolved clumps in a $10^{-5} \sr$ 
solid angle along the direction $\psi$, for each substructure model considered. The result is shown in  
Fig.~\ref{fig3}. In the same figure, the solid thick line corresponds to the contribution from the MW smooth halo component described in Sec.\ref{sec:diffgal} , which is computed according to Eq.\ref{singlehalophicosmo}, with the distance to the observer $\lambda = \rsun$.

We summarize the properties of the smooth subhalo contribution in
Table 2. In the second and third column we show, 
for each model, the contribution to $\Phi^{\rm cosmo}$ 
in units of $\GeV^2 \cm^{-6} \kpc \sr$ towards the Galactic 
center ($\psi=0^\circ$) and the angle $\psi_d$ beyond which the 
smooth subhalo contribution starts dominating over 
the MW halo foreground.
In the fourth column we show the boost factor for each model,
computed as the ratio of the integral over the MW volume of the density squared including subhaloes to the same integral for the smooth MW only:
\begin{equation}
b = \frac{\int_{MW} dV ( \rho_{MW smooth}^2 + \rho_{sh}^2 )}{\int_{MW} dV \rho_{MW smooth}^2}
\label{boostfactor}
\end{equation}

\begin{figure}
\epsfig{file=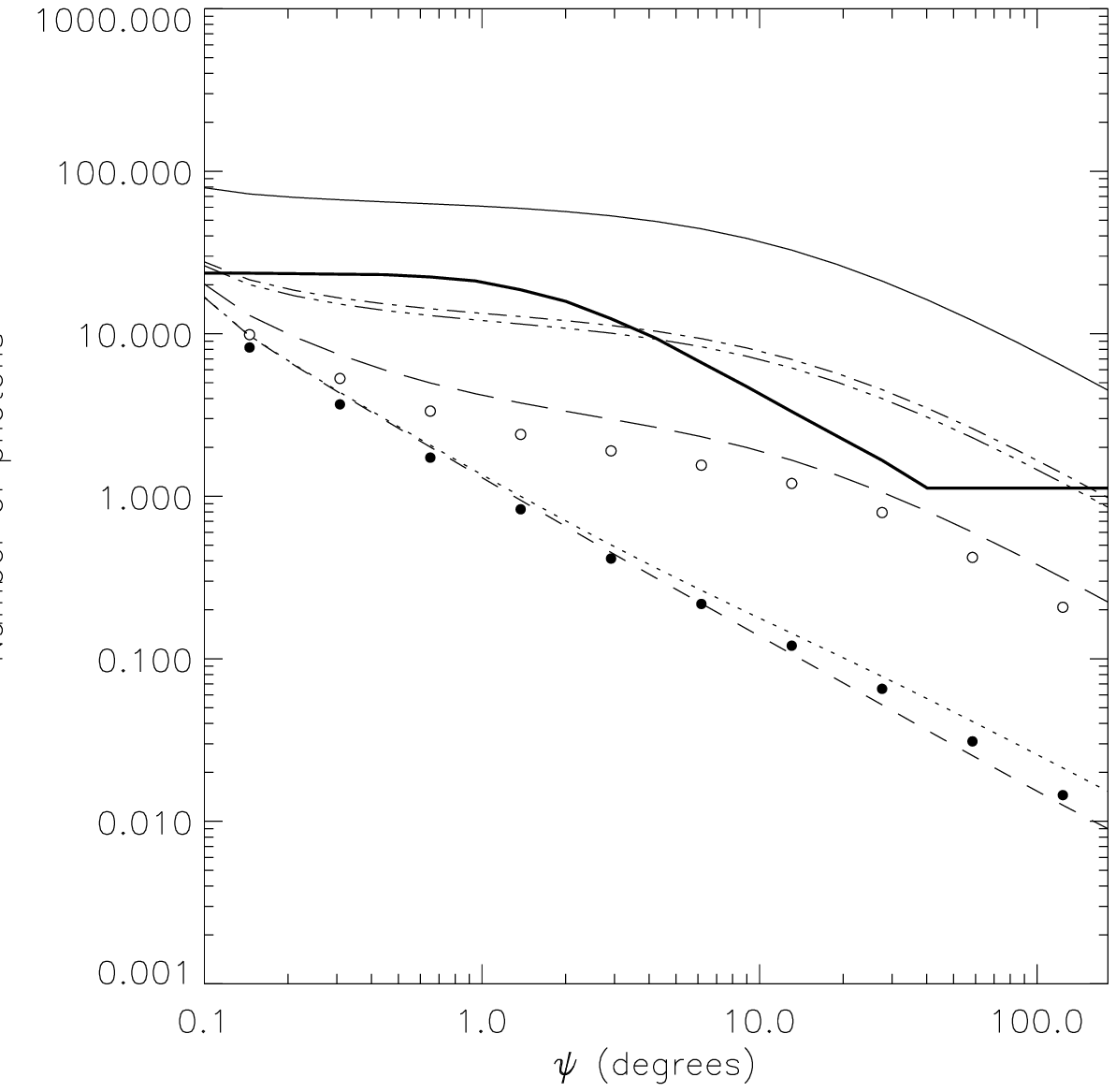,width=0.65\textwidth}
\caption{Number of photons above 3 GeV, in 1 year in a solid angle of $10^{-5} \sr$, as a function of the angle $\psi$ from the GC.
From top to bottom, the lines correspond to the \Bd, \Bc, \Eb, \Bb, \Ba and \Ea model. Empty (filled) circles show the \Bzref (\Bref) model.
$\sigma_{\rm ann} v = 3 \times 10^{-26} \cm^3 \sec^{-1}$, $m_\chi = 40 \GeV$ and $BR_{b \bar b} = 100 \%$ have been used, corresponding to the best value $\Phi_{PP} = 2.6 \times 10^{-9} \cm^4 \kpc^{-1} \GeV^{-2} \sec^{-1} \sr^{-1}$.
The solid thick line shows the EGRET diffuse expected Galactic and extragalactic background computed along $l=0$. The mass function slope is $\alpha=1$.
}
\label{fig5}
\end{figure}
\begin{table}
\begin{center}
\begin{tabular}{|c|c|c|c|c|}
\hline 
  & \multicolumn{4}{|c|}{Results for subhalo models} \\ \hline
Model & $\Phi^{\rm cosmo}_{0}$ & $\psi_d$ & b & $\Phi^{PP}_{-9}$  \\ \hline
\Bref & $1.4 \times 10^{-4}$  & 7.5 & 8&  2.6  \\ \hline
\Ba & $2.0 \times 10^{-4}$  &  7.5 & 10 & $2.6 $  \\ \hline
\Bb & $3.6 \times 10^{-3}$ & 0.4 & 150 & $2.6 $  \\ \hline
\Bzref & $2.1 \times 10^{-3}$  & 0.6 & 120 & 2.6  \\ \hline
\Bc & $1.6 \times 10^{-2}$ &  0.1 & 750 & $0.9 $  \\ \hline
\Bd & $7.7 \times 10^{-2}$ &  0.0 & 3300 & $0.2 $  \\ \hline
\Ea & $1.2 \times 10^{-4}$ &  12.0 & 6 & $2.6 $  \\ \hline
\Eb & $1.4 \times 10^{-2}$ &  0.16 & 660 & $1.0 $  \\ \hline
\end{tabular}
\label{tab2} 
\caption{Results for the halo models. Column 1: Model name. Column 2: $\Phi^{\rm cosmo}$ [$\GeV^2 \cm^{-6} \kpc \sr$] toward the Galactic Center. 
Column 3: Angle at which the subhalo diffuse contribution dominates over the MW smooth foreground [degrees]. Column 4:
boost factors. Column 5: $\Phi^{PP}_{-9}$ [ $10^{-9} \cm^4 \kpc^{-1} \GeV^{-2} \sec^{-1} \sr^{-1}$]. The value for the {\it z0} and the \Bzref models correspond to our best case particle physics scenario. Values for the {\it zc} models except the \Bzref are normalized to EGRET data.}
\end{center}
\end{table}

To set the two remaining parameters that determine the intensity of the annihilation flux,
namely the particle mass
$m_\chi$ and the annihilation cross section 
$\sigma_{\rm ann}v$ we adopt the most optimistic combination allowed by the 
SUSY and UED models shown in Fig.~\ref{sigmavmchi} that do not not exceed the current 
EGRET upper limits for the annihilation flux above 3 GeV and within 
a  solid angle of $10^{-5} \sr$.
The latter receives contribution from two distinct components:
the first one is of Galactic origin,
dominates for $\psi < 40^{\circ}$ and is characterized by a power-law
photon spectrum, that leads, upon extrapolation at high energies (Bergstr\"om \etal 1998)
to the following parametrization
\begin{equation}
\frac{d \phi^{\rm gal-\gamma}_{\rm diffuse}}{d\Omega dE}=
 N_0(l,b) \;10^{-6}\; E_{\gamma}^{-2.7} \frac{\gamma}{\cm^2 \sec \sr \GeV},
\label{dndegal}
\end{equation}
where $l$ and $b$ are the galactic latitude and longitude.
The normalization factor $N_0$ depends only on the interstellar matter distribution, and is
modeled as in  Bergstrom \etal 1998.

The second one is extragalactic,
dominates at $\psi > 40^{\circ}$, and for its photon spectrum
we use an extrapolation from low energy EGRET data, following Sreekumar \etal 1998:
\begin{equation}
\frac{d \phi^{\rm extra-\gamma}_{\rm diffuse}}{d\Omega dE} 
= 1.38 \times 10^{-6} E^{-2.1} 
\frac{\gamma}{\cm^2 \sec \sr \GeV}. 
\label{gammas}
\end{equation}

The thick line in Fig.~\ref{fig5} shows the EGRET photon flux as a function of  $\psi$. The EGRET flux is computed according to Eqs. ~\ref{dndegal}
and ~\ref{gammas} extrapolated above 3 GeV, within a field of view of  $10^{-5}$ sr in  1 year of
observation. The flux is computed  along $l=0$, that is away from the Galactic plane, 
where the EGRET flux is minimum.
The Galactic and extragalactic contributions
are clearly visible.
The other curves show the predictions for all models in Table 1, obtained when  
using our best case particle physics scenario described in Sec.~\ref{fluxpp}
that corresponds to $\Phi^{PP}=2.6\times 10^{-9} \cm^4 \kpc^{-1} \GeV^{-2} \sec^{-1} \sr^{-1}$.
All the values of $\sigma_{\rm ann} v$ and $m_\chi$ that would result in the same  $\Phi^{PP}$ 
are represented with a solid line in Fig.\ref{sigmavmchi}.

With this choice of parameters all models flagged with {\it zc} (but the \Bzref), for which the halo properties are 
computed at the collapse redshift,
exceed experimental data close to the GC and 
for $\psi > 40^{\circ}$, where the EGRET flux is assumed to have an extragalactic origin.
We do not regard the mismatch at small angles as significant, because of 
the limited angular resolution of the original EGRET data.
On the contrary, 
 we decrease the values of  $\Phi^{PP}$ to bring the {\it zc}-models into agreement with data at large 
 angles from the GC.
The values of $\Phi^{PP}$ adopted in each model are  listed in the last column of Table 2. 
We note that a smaller $\Phi^{PP}$  corresponds to assuming a larger particle mass or a smaller 
cross section. 
The $[\sigma_{\rm ann} v, m_\chi]$ phase space parameter allowed for SUSY (circles) or UED (dotted line) models is shown in Fig.~\ref{sigmavmchi}. 
In the same figure the three dashed and dot-dashed lines correspond to the EGRET constrained values of $\Phi^{PP}$ for the {\it zc} models. The particle physics models above the corresponding line are thus excluded by EGRET data.

\begin{figure*}
\hspace{2truecm}
\epsfig{file=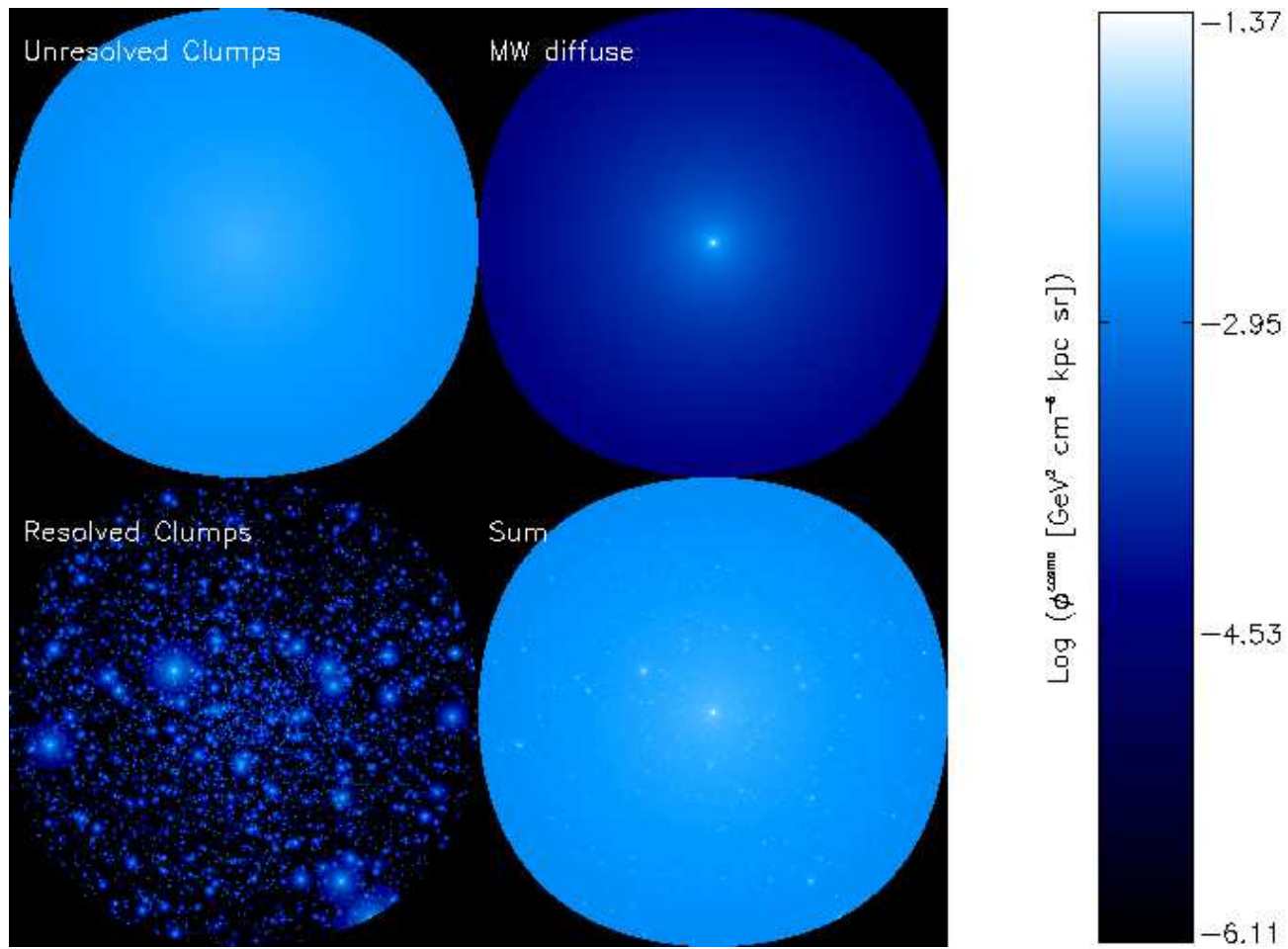,width=\textwidth}
\caption{Map of $\Phi^{\rm cosmo}$ (proportional to the annihilation signal) for the \Bzref model, in a cone of $50^\circ$ around the Galactic Center, as seen from the position of the Sun. Upper left: smooth subhalo contribution from unresolved haloes. Upper right: MW smooth contribution. Lower left: contribution from resolved haloes. Lower right: sum of the three contributions.
}
\label{fig6}
\end{figure*}

\subsection{Annihilation flux from individual clumps}
Besides the diffuse signal produced by annihilation in
both the subhalo population and the  smooth MW component,
we consider here the contribution from individual subhaloes, that 
we regard as Poisson fluctuation of the underlying mean distribution 
of subgalactic haloes that could be detected as isolated structures.
To estimate their flux
we consider 10 independent Monte Carlo realizations of the closest and brightest subhaloes, in a 
cone of $\sim 50^\circ$ pointing toward the Galactic Center.  
To do this we generate, for each mass decade, the positions of those subhaloes that have 
$\Phi^{\rm cosmo} > \langle \Phi^{\rm cosmo}_{B_{z0}}(\psi = 50^\circ) \rangle \sim 5 \times 10^{-5} \GeV^2 \cm^{-6} \kpc \sr$ , where the brackets indicate
the mean annihilation flux. If  $N<100$ such objects are found, then we still include the
remaining $100-N$ nearest subhaloes in that mass range.

Adding contribution from an increasing number of 
individual haloes monotonically increases the chance of subhalo detection within the angular resolution element
of the detector. 
To check whether our procedure is robust and the number of detectable haloes 
has converged
we reduced the number of Monte Carlo-generated haloes by 70 \% 
and found that probability of subhalo detection indeed remains constant.

\begin{figure*}
\hspace{1truecm}
\epsfig{file=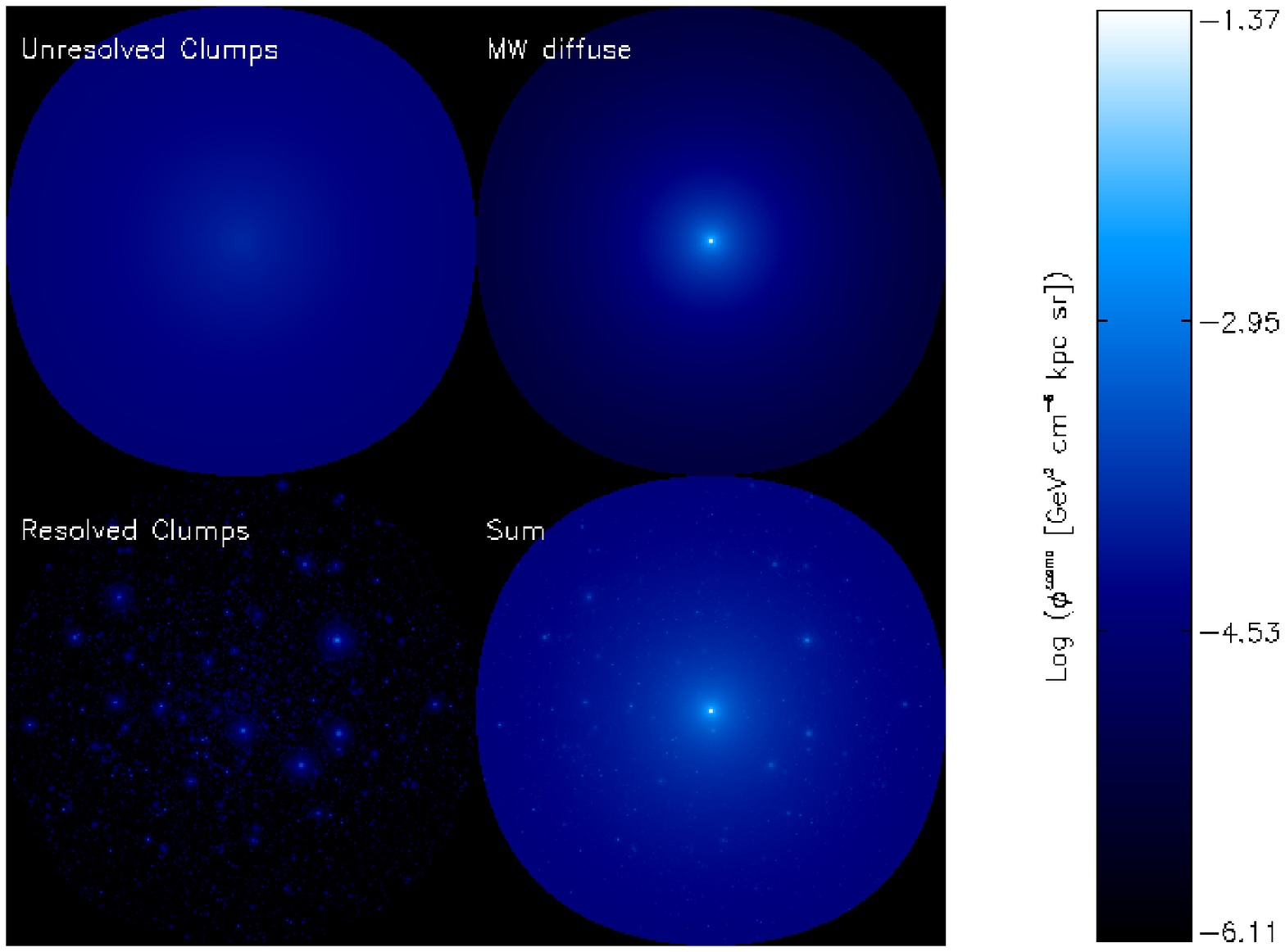,width=\textwidth}
\caption{
Same as in Fig. \ref{fig6} for the \Ea model.
}
\label{fig7}
\end{figure*}

To summarize, for each model, the total contribution to $\Phi^{\rm cosmo}$ is given by the sum of three terms:
the diffuse contribution coming from unresolved haloes, corresponding to the mean contribution of the clumpy component
computed with Eq.\ref{smoothphicosmo}, the contribution of the diffuse smooth Galactic component
and that of individual nearby subhaloes, both computed using Eq.\ref{singlehalophicosmo}.
Figs.\ref{fig6} and \ref{fig7} show the three contributions to $\Phi^{\rm cosmo}$ for the models \Bzref and \Ea, respectively. In each figure, the contribution from unresolved clumps is shown in the upper left panel, the one  from the diffuse MW in the upper right, the one from resolved clumps in the lower left, and the sum of all contributions in the lower left panel. The smooth MW halo contribution falls rapidly with the distance from the GC, while the diffuse subhalo contribution keeps a high value even at large angular distances. The single halo contribution is almost completely hidden by the overwhelming 
diffuse foreground, while it is nicely resolved as a standalone component.

The same procedure described in this section for a cone pointing to the GC has been repeated for two other regions: a cone pointing to the Galactic anticenter and a cone to the Galactic pole $b = 90^\circ$. 
The purpose is to compute the annihilation flux and to evaluate the number of detectable subhalos over
the whole sky. \\

\section{Prospects for detection}
\label{sec:detection}

In absence of strong features in the annihilation spectrum, the best chances to 
detect the annihilation signal within our Galaxy is to observe some excess on the $\gamma$-ray sky
either due to diffuse emission or to resolved sources that have no astrophysical 
counterpart.
However, the requirements for signal detection are different in the two cases since 
the smooth annihilation flux, that contributes to the signal in the first case,
adds to the noise in the second.

To determine the probability of halo detection, 
we consider a 1 year effective exposure time performed with a GLAST-like satellite.
We note that, given the 2.4 sr field of view and the all-sky survey mode of GLAST, such an exposure will
be achieved in about 5 years of actual observation time.

The prospects for detecting $\gamma$-rays from DM annihilations are evaluated by
comparing the number $n_\gamma$ of expected signal photons to  the
fluctuations of background events $n_{\rm bkg}$. To this purpose we
define the sensitivity $\sigma$ as:
\begin{eqnarray}
\sigma &\equiv& \frac{n_{\gamma}}{\sqrt{n_{\rm bkg}}}\\ &=&
\sqrt{T_\delta} \epsilon_{\Delta \Omega}
\frac{\int A^{\rm eff}_\gamma (E,\theta_i) [d\phi^{\rm signal}_\gamma/dE
d\Omega] dE d\Omega}{\sqrt{ \int \sum_{\rm bkg} A^{\rm eff}_{\rm
bkg}(E,\theta_i) [d\phi_{bkg}/dEd\Omega] dE d\Omega}} \nonumber
\label{sensitivity}
\end{eqnarray}
where $T_\delta$ defines the effective observation time and
$\phi_{bkg}$ is the background flux. 
The quantity $\epsilon_{\Delta \Omega}$ is the fraction of signal
events within the optimal solid angle $\Delta \Omega$ corresponding to
the angular resolution of the instrument and is usually $\sim 0.7$. 
We set it equal to 1 to get the most optimistic values. 
The effective detection
area $A^{\rm eff}$ for electromagnetic or hadronic particles
is defined as the detection efficiency times the geometrical
detection area.
In the following we make the realistic assumption that all hadronic
particles will be identified, so that the background will be composed by
photons only.
We also assume that all photons will be correctly identified, which is somehow optimistic,
since there will be a small amount (a few percent) of irreducible background due, e.g.,
to the backsplash of high energy photons. \\
We use $A^{\rm eff} = 10^4 \cm^2$, independent
from the energy $E$ and the incidence angle $\theta_i$, and an angular resolution of 
 $0.1^{\circ}$. Both these values are rather optimistic, since the expected GLAST angular resolution 
approaches $0.1^{\circ}$ only at about 20 GeV, while the on-axis effective area 
is quoted to be maximum $\sim 8 \times 10^3 \cm^2$ above 1 GeV and decreases by  
$\sim$20\% for an incidence angle of $20^\circ$.

As anticipated, different annihilation signals need to be compared with different background noises.
For the detection of the diffuse annihilation flux the background is contributed 
both by Galactic  (Eq. ~\ref{dndegal}) and extragalactic astrophysical sources (Eq. \ref{gammas})
measured by EGRET. 
In the case of individual subhaloes, the annihilation photons produced in the smooth Galactic halo
and in the unresolved clumpy component contribute to the  background rather than to the signal.

\subsection{Sensitivity to diffuse emission}
We first study the sensitivity $\sigma$ of such a GLAST-like observatory to the annihilation flux from
the smooth DM profile and from the diffuse contribution of unresolved subhaloes. 
Both signals are computed above $3 \GeV$, and 
the astrophysical background noise is obtained from Eqs.\ref{dndegal} and \ref{gammas} specified  along $l=0$.

\begin{figure}
\epsfig{file=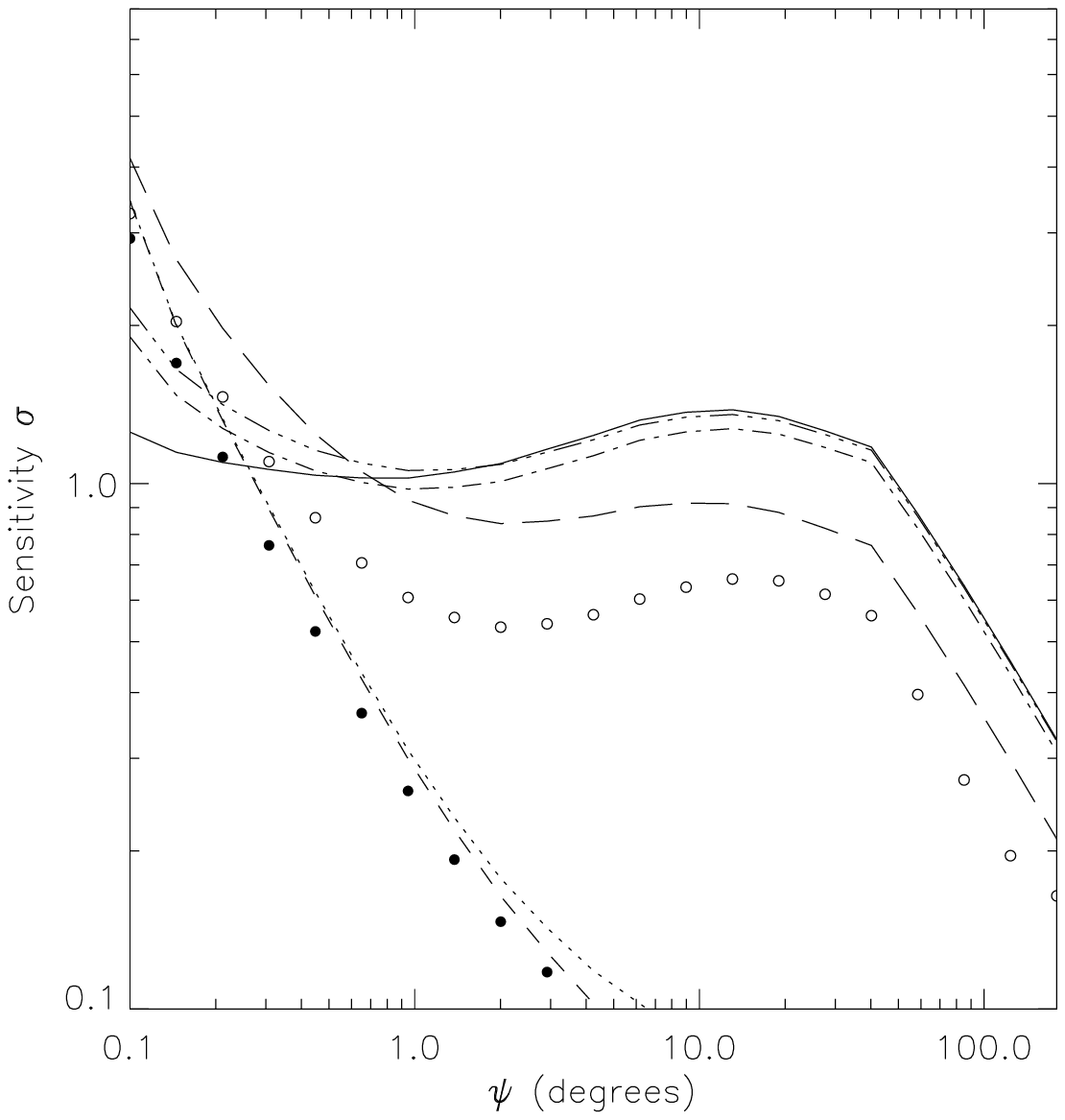,width=0.65\textwidth}
\caption{
Statistical significance, as a function of the angle from the GC $\psi$, for the detection 
of the DM annihilation flux from diffuse subhaloes plus the MW smooth 
component (along $l=0$), for the different models explored:
 \Ba (dotted line ), \Ea (short dashed), \Bref (filled circles), \Bzref (empty circles), \Bb (long dashed), \Bc (long dot-dashed), \Eb (dot-dashed), \Bd (solid). The mass function slope is $\alpha=1$.
 We refer to Table 2 for the values of $\Phi_{PP}$ used in this figure.
}
\label{fig8}
\end{figure}

The result is shown in Fig.\ref{fig8}, where we plot the statistical significance of the detection as a function of 
$\psi$ for each of the models listed in  Table \ref{tab1}.
1 $\sigma$ detections of the annihilation signal is expected at  $\psi < 40^\circ$ for all the models labelled {\it zc}.
The chances of observing the diffuse annihilation flux are significantly higher in the direction of the 
Galactic Center along which models labelled {\it z0} predict a signal detectability 
as high as 5 $\sigma$. Yet, these predictions should be taken with much care since the measured astrophysical
$\gamma$-ray flux above $3 \GeV$ in the direction of the GC, which constitutes the background, 
is known with large uncertainties.

\subsection{Detection of individual haloes}

Subhaloes can also be detected through the annihilation flux produced by individual, nearby clumps that 
would appear as bright, possibly extended, sources, as shown in the bottom right panels
 of Figs. \ref{fig6} and \ref{fig7}.
In this case the signal is  produced within the individual haloes of our Monte Carlo realizations,
 while the background is contributed by the smooth astrophysical background plus the diffuse annihilation
flux produced by the Galaxy and its subhaloes. \\
For each halo in the 10 Monte Carlo realizations, and for each virial concentration model, we
 
\begin{itemize}
\item assign to it an arbitrary concentration parameter $c(M)$;
\item calculate the annihilation signal;
\item find the value of the concentration parameter that guarantees a 5 $\sigma$ detection in 1 year  exposure time,  $c_{5 \sigma}(M)$;
\item identify the probability of detection of the clump with the probability $P(>c_{5 \sigma})$ that such a clump has a concentration as high as  $c_{5 \sigma}(M)$, assuming the lognormal distribution described in Eq.\ref{pcvir}.
\end{itemize}

The total number of detectable subhaloes is then simply given by $\sum_i P_i(>c_{5 \sigma})$, 
where the sum is performed over all haloes in the realization. 
Results are obtained by averaging over all over the 10  Monte Carlo realizations 
and the procedure is repeated for all models listed in Table 1.

\begin{figure}
\epsfig{file=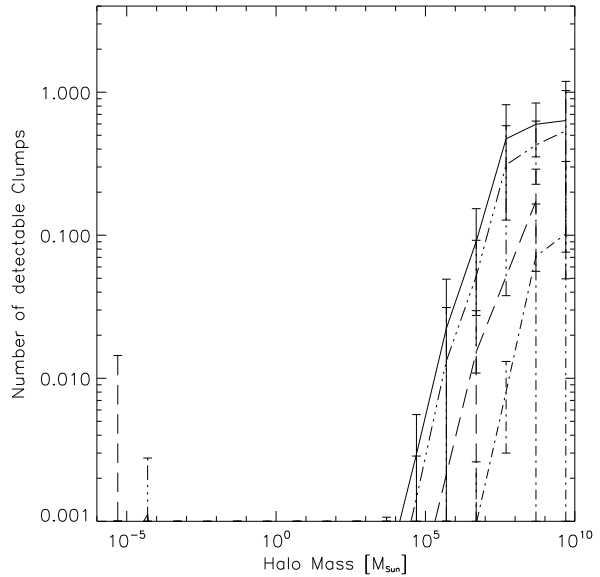,width=0.65\textwidth}
\caption{Number of events detectable in 1 year a 5 $\sigma$ with a GLAST-like experiment in a 50 degrees cone towards the GC for the models \Ea (dot-dashed), \Ba (solid), \Bb (dashed) and \Bref (long dot-dashed), assuming the lognormal distribution for the concentration parameter. We refer to Table 2 for the used values of $\Phi_{PP}$. The mass function slope is $\alpha=1$.
}
\label{fig10}
\end{figure}

\begin{figure}
\epsfig{file=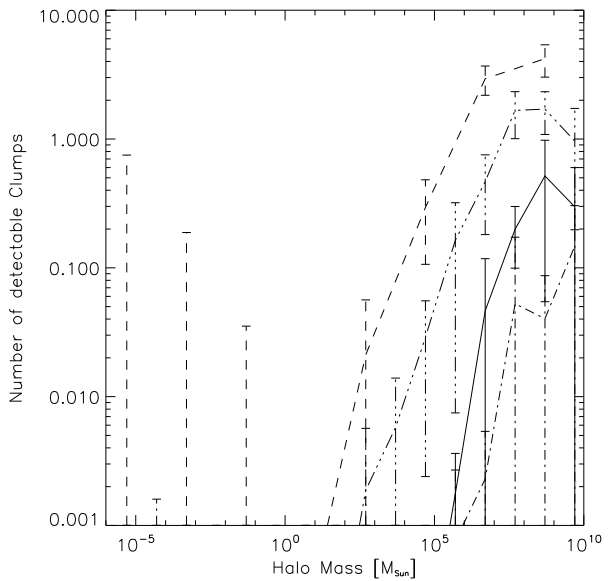,width=0.65\textwidth}
\caption{Number of events detectable in 1 year a 5 $\sigma$ with a GLAST-like experiment in a 50 degrees cone towards the GC for
the models \Bzref (dashed) \Eb (solid), \Bc (long dot-dashed) and \Bd (dot-dashed), assuming the lognormal distribution for the concentration parameter. We refer to Table 2 for the used values of $\Phi_{PP}$. The mass function slope is $\alpha=1$.
}
\label{fig11}
\end{figure}

The number of haloes that can be detected in 1 year with a significance above
5 $\sigma$ in  cone of view with angular opening of $50^\circ$ towards the GC
is shown in Fig.\ref{fig10} for the {\it z0} models and in Fig. \ref{fig11} for the {\it zc} models, as a function of the subhaloes mass. 
Had we assumed a deterministic relation for $c(M)$ the number of events would have decreased by a factor $\sim 2$.

At higher latitudes the number of haloes indeed reduces, but this is compensated by the lower foreground given by the smooth subhalo component.
The maximum number of detectable events is obtained toward $b = 90$ where these two effects interplay in a most favourable way for the detection. 

Table 3 lists the number of haloes that can be detected with a significance larger than 5 $\sigma$ in a cone of $50^\circ$ around the Galactic Center (first column), the Galactic pole (second column) and the Galactic anticenter (third column), for each model, for our reference mass function slope $\alpha=1$ and the $\Phi_{PP}$ values listed in Table 2. 
Hereafter, each error is the standard deviation obtained averaging over the 10 MC representations. 
The effect of decreasing the number of haloes far from the GC is compensated by the lower foreground due to the diffuse subhalo contribution to the annihilation flux. The best compromise is found around the Galactic poles.

\begin{table}
\begin{center}
\begin{tabular}{|c|c|c|c|}
\hline 
  & \multicolumn{3}{|c|}{Number of detectable haloes ($\alpha=1$)} \\ \hline
Model & $N^{5 \sigma}_{GC}$   &  $N^{5 \sigma}_{90}$ &  $N^{5 \sigma}_{180}$ \\ \hline
\Bref & $0.65 \pm 0.45$ & $0.85 \pm 0.43$ & $0.59 \pm 0.30$   \\ \hline
\Ba & $0.65 \pm 0.45$ & $0.84 \pm 0.43$  & $0.59 \pm 0.30$ \\ \hline
\Bb & $0.46 \pm 0.34$ & $0.60 \pm 0.34$ & $0.50 \pm 0.27$ \\ \hline
\Bzref &  $16.16 \pm 2.60$ & $23.24 \pm 2.28$ &  $18.55 \pm 1.72$ \\ \hline
\Bc & $1.74 \pm 0.92$ &  $2.40 \pm 0.82$ & $2.15 \pm 0.65$ \\ \hline
\Bd & $0.05 \pm 0.05$ & $0.07 \pm 0.08$ & $0.08 \pm 0.08$  \\ \hline
\Ea & $0.06 \pm 0.12$ & $0.07 \pm 0.10$ & $0.04 \pm 0.07$  \\ \hline
\Eb & $0.29 \pm 0.40$ & $0.49 \pm 0.37$ & $0.46 \pm 0.30$ \\ \hline
\end{tabular}
\label{tab3} 
\caption{Number of haloes detectable, at 5 $\sigma$ in 1 year of effective observation with a GLAST-like satellite, in a $50^\circ$ f.o.v. cone towards the GC (column 1),  the Galactic pole (column 2) and the anticenter (column 3). 
The subhaloes mass function slope is $\alpha = 1$.
We refer to Table 2 for the values of $\Phi_{PP}$ used in this table.
}
\end{center}
\end{table}  

In Table 4 we show the total number of haloes that can be detected with a GLAST-like satellite in the whole sky with a significance larger than 5 $\sigma$, with a mass function slope $\alpha = 1$ (first column) and $\Phi_{PP}$ values listed in Table 2. 

The remaining two columns show the effect of adopting a  mass function with 
power-law index $\alpha = 0.95$ (second column) and $\alpha = 0.9$ (third column).
Adopting a shallower mass function increases in most cases the number of detectable subhalos. However, the magnitude of the effect, that results from  the lowering of the unresolved background, whose main contribution is given by small sub-haloes, depends on the model explored. 

As expected, we can observe how this effect is larger for those models whose overall contribution to $\Phi^{cosmo}$ is larger, that is for those models whose concentration parameters have been computed at the collapse redshift. 
The effect is reduced for the other models, as well as for the \Bzref one, for the following reason:
when using $\alpha=1$, the $\Phi^{PP}$ value for the $zc$ models (but the \Bzref one) has been decreased in order to respect the EGRET EGB limit, while when using $\alpha=0.9$ all models fulfill the EGRET EGB constraint, and we can use our best case $\Phi_{PP}$. This is also true for $\alpha=0.95$, except for the \Bd model, where we have to use  $\Phi^{PP} = 0.84 \cm^4 \kpc^{-1} \GeV^{-2} \sec^{-1} \sr^{-1}$.
In fact, the $z0$ and the \Bzref model experience just a minor increase (compatible within the error bars) of the number of detectable halos; the \Bc and \Eb models reach this stability for $\alpha \le 0.95$, as soon as their $\Phi^{PP}$ allowed value gets our best value; the \Bd model keeps on showing a large effect when changing mass function slope, because it is allowed to have the best value $\Phi^{PP}$ only when $\alpha=0.9$.

In the most optimistic \Bzref model, we expect that a GLAST-like experiment could detect $\sim 120-130$  subhaloes with masses above
$10^{5} M\odot$ over all sky, for all the mass function slopes considered in this analysis. 
In all the models whose concentration parameters are computed at $z=0$ the number of detectable events is compatible with zero within the errors, whatever slope is used.
Accordingly to the afore-mentioned discussion, the effect of changing the mass function slope is dramatic in the \Bc (\Bd) model, for which the total number of events ranges from $\sim 10$ ($\sim0$) for $\alpha=1$ to $\sim 120$ ($\sim 100$) for $\alpha=0.9$. A large effect ($\sim 0$ to $\sim 30$) is observed in the \Eb model too.

\begin{table}
\begin{center}
\begin{tabular}{|c|c|c|c|}
\hline 
  & \multicolumn{3}{|c|}{Total number of detectable haloes} \\ \hline
Model & $N^{5 \sigma}_{tot}$ ($\alpha=1)$ &  $N^{5 \sigma}_{tot}$ ($\alpha=0.95)$ &  $N^{5 \sigma}_{tot}$ ($\alpha=0.9)$ \\ \hline
\Bref & $4.30 \pm 4.00$ & $3.62 \pm 3.30$ & $3.51 \pm 2.11$ \\ \hline
\Ba &  $4.26 \pm 3.97$ & $3.61 \pm 3.30$& $3.50 \pm 2.13$ \\ \hline
\Bb & $3.12 \pm 3.09$ & $3.30 \pm 3.17$& $3.43 \pm 2.04$ \\ \hline
\Bzref &  $118.36 \pm 24.96$ & $132.89 \pm 30.15$ & $125.03 \pm 20.06$\\ \hline
\Bc & $12.53 \pm 8.67$ & $104.23 \pm 24.78$ & $119.04 \pm 19.77$ \\ \hline
\Bd &  $0.39 \pm 0.56$ & $10.55 \pm 6.36$ & $96.34 \pm 18.66$\\ \hline
\Ea &  $0.33 \pm 0.89$ & $0.67 \pm 1.58$ & $0.34 \pm 0.50$\\ \hline
\Eb & $2.50 \pm 4.48$ &$23.43 \pm 10.17$ &$30.40 \pm 10.31$ \\ \hline
\end{tabular}
\label{tab4} 
\caption{Total number of haloes detectable over the whole sky, at 5 $\sigma$ in 1 year of effective observation with a GLAST-like satellite, for a mass function slope $\alpha = 1$ (column 1), $\alpha = 0.95$ (column 
2) and $\alpha = 0.9$ (column 3). We refer to Table 2 for the values of $\Phi_{PP}$ used in this table.}
\end{center}
\end{table}

\section{Discussion and Conclusions}
\label{sec:conclusions}

\begin{figure}
\epsfig{file=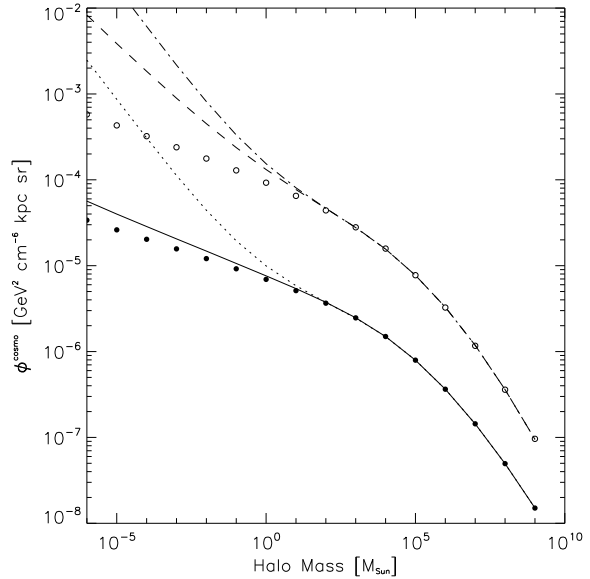,width=0.65\textwidth}
\caption{Contribution to $\Phi_{cosmo}$ from substructures of mass M integrated over the mass decade, for the different $B$ models explored and computed toward the Galactic Center. Solid and empty circles correspond to the \Bref and \Bzref models respectively. Lines show the \Bd (dot-dashed), \Bc (dashed), \Bb (dotted) and \Ba (solid) models. The mass function slope is $\alpha=1$.}
\label{phivsmass}
\end{figure}

The prospects for detecting $\gamma$-rays from
the annihilation of DM particles in substructures of the MW have been investigated by a number of authors (e.g. Stoehr \etal 2003, 
Pieri \& Branchini 2004, Koushiappas \etal 2004, Oda \etal 2005, Pieri \etal 2005). In this work 
we confirm that substructures can provide a significant contribution to the expected Galactic annihilation
signal, although the actual enhancement depends on the assumptions made on the clump properties, which are affected by large uncertainties. Indeed, given the assumed substructure mass function 
$dN/dM \propto M^{-2}$, the contribution to the total $\gamma$-ray flux  by subhaloes of different
masses depends on the annihilation signal produced within each clump which is dictated by
the internal structure. Numerical experiments have shown that the total annihilation 
signal is dominated by the highest mass subhaloes both in a Galactic halo at z=0
(Stoehr \etal 2003) and in 0.1 $\msun$ halo host at z=75 (Diemand \etal 2006).
However, the recent results of the high resolution 'Via Lactea' simulation 
(Diemand et al 2007a) indicate that the annihilation luminosity is approximately constant per decade of substructure mass while 
analytical calculations (Colafrancesco \etal 2006) tend to find that the signal is dominated by small mass subhaloes. Indeed this is also the case with our model predictions.
Fig.~\ref{phivsmass} shows the expected contribution of the unresolved haloes to the total annihilation flux as a function of the subhalo mass, integrated on each mass decade.
In all the models explored the annihilation signal
is dominated by the smallest clumps, as a result of the decrease of the  virial concentration
with the subhalo mass, as  shown in Fig. \ref{fig2}. 
Under optmistic assumptions on the particle physics parameters of DM particles, a GLAST-like experiment 
might detect such a DM annihilation flux, but only in a few pixels around the Galactic Center.

It should be noticed, however, that estimates of the annihilation signal from 
the Galactic Center  
are affected by the poor knowledge of the DM profile in the innermost regions
of the Galaxy, which is usually obtained by extrapolating over many orders of 
magnitude the results of numerical simulations. The presence of a Supermassive Black Hole at the
center of the Galaxy makes things even more complicated, as it may significantly
affect the distribution of DM within its radius of gravitational influence ,
leading to the formation of an overdensity called "spike" (Gondolo and Silk, 1999).
Spikes require however rather fine-tuned conditions to form (Ullio et al. 2002)
and any overdensity is in any case severely suppressed by the interaction
with stars and DM self-annihilations (Merritt et al. 2002, Bertone \& Merritt 2005).

In alternative, one could look for an annihilation signal from individual 
DM substructures, such as dwarf galaxies or even smaller, 
'baryon-less', clumps.
We have shown that, depending on the assumptions made on the properties of 
clumps, only large haloes with $M > 10^5 \msun$ can be detected
and identified with a GLAST-like experiment, which is consistent with the 
analyses of Stoher \etal 2003 and Koushiappas \etal 2004. The number of detectable haloes ranges from 0 to more than a hundred, depending on the model.
Adopting a shallower subhalo mass function increases the 
number of detectable subhaloes in those models in which the diffuse annihilation signal is dominated by the unresolved, low mass haloes.

In any case, scenarios leading to a large number of detectable
small-scale clumps appear to be 
severely constrained by the $\gamma$-ray background measured by EGRET. 

In particular, the model of  Koushiappas 2006 is similar to our \Bc
model, as far the cosmological term is considered, while the particle physics 
contribution corresponds to our best case scenario $\Phi^{PP}_{B_{z_0}}$. 
Although nearby haloes would be bright, and observable, in this case, 
we have shown that the associated diffuse emission produced by all the 
remaining, unresolved, clumps in the Milky Way, would far exceed the 
$\gamma$-ray background measured by EGRET. The chances to detect the proper 
motion of clumps are thus very low, as the lowest mass detectable subhaloes, 
(M=$10^5 \msun$), are typically found at a distance greater than
 $0.5 \kpc$, leading to a proper motion 
less than  $\sim 0.1' \ {\rm yr}^{-1}$, well below the GLAST angular 
resolution of a few arcminutes.

We have made use of simplified and extreme scenarios for the subhaloes concentration parameter models. More accurate scenarios, though not supported by numerical simulations for small mass haloes, could lead to different diffuse foreground levels and to both more or fewer detectable haloes.  

One may wonder why we preferentially expect to individually detect the more 
massive subhaloes, 
while the unresolved annihilation signal is mainly contributed by small
mass clumps. The reason is that the volume over which 
individual haloes can be detected decreases rapidly with the halo mass. 
To see this, let us consider the maximum distance $D_{\rm MAX}$
at which a clump can be detected.
This distance depends on the halo luminosity which, in turns, 
depends  on the halo mass and concentration. For a NFW profile: $D_{\rm MAX} \propto
M^{0.5} c(M)^{1.5}$.  On the other hand, as discussed e.g. by Koushiappas
2006,  given a subhalo mass function $dN/dM\propto M^{-2}$,
the number of detectable haloes  per mass decade is:
$ dN/dLog(M) \propto D_{\rm MAX}^3 M^{-1} \propto M^{0.5} c(M)^{4.5}$.
Assuming a simple scale-free virial concentration  $c(M)\propto M^{-\gamma}$, we see that 
if $\gamma>\gamma_{th}=-1/9$, then the  number of detectable haloes per mass decade halo
indeed increases with the subhalo mass.
The  $c(M)$ relation for some of our models is shown in Fig. 2 along with the 
 $\gamma_{th} =-1/9$ reference slope (thick dashed line). 
 Models  \Ba and \Ea, that have $\gamma>\gamma_{th}$ do indeed 
 predict that the probability of subhalo detection increases with the mass
 (see Fig. 9). On the other hand, the slope of the $c(M)$ relation for $M<10^2 \msun$
 is sligthly steeper than $\gamma_{th}$. Therefore we expect a bimodal probability that peaks at
 high masses with a secondary maximum  for the smallest subhaloes. Indeed, this is 
 what we observe in Figure 9.

In conclusion, we have studied the prospects for indirect detection of Dark Matter in MW subhaloes with a GLAST-like satellite. 
We have chosen 8 different models for the concentration parameter, which span the phase space the theoretical uncertainties on the Dark Matter halo properties, as well as 3 different values for the subhaloes mass function slope. For each model, we have computed the diffuse emission from unresolved subhaloes, as well as the $\gamma$-ray flux from individual, nearby haloes.  

We found that for models with concentration parameter computed at $z=0$ the detection of individual haloes appears challenging, while the diffuse emission from unresolved clumps dominates the MW smooth emission for sky directions  $>1$ degree off the GC. 

In the case of $\alpha=1$, in all the $zc$ models except the \Bzref, the diffuse emission from unresolved clumps exceeds the EGRET constraints in a portion of the DM parameter space relevant for SUSY models, as shown in Fig. \ref{sigmavmchi}. Adopting DM models compatible with the EGRET data, one may still hope to detect individual haloes, like e.g. in the \Bc model.

The \Bzref is our best case model for all the values of the mass function slope, though it should be stressed that it is not supported by numerical experiments but only by theoretical considerations. 
The success of the \Bzref model is due to the fact that the diffuse emission expected from subhaloes is dominated by small mass haloes while the large mass haloes are most favourably detected as spare sources. A functional form for the concentration parameter which flattens at low masses, as it is the \Bzref one, will in fact decrease the diffuse emission, thus allowing a larger value for $\Phi_{PP}$ and consequently increasing the chances of detection for large mass haloes.

In general, adopting the most optimistic set of parameters for the DM particle compatible with EGRET (see Table 2), the number of detectable subhaloes over all sky (at 5$\sigma$, with a GLAST-like experiment and a 1-year exposure time), for the 8 x 3 models we have studied, ranges between $0$ and $120$. 

Yet, it should be noticed that the numbers listed in Table 2 are obtained with a very optimistic value for the Particle Physics involved in the process.
If we assume a more realistic model for the Dark Matter particle ($\m_\chi = 100 \GeV \ , \sigma v = 10^{-26} \cm^3 \sec^{-1}$) instead, we find that at most only 
a handful of detectable haloes are found (at most a handful over all sky for the most optimistic \Bzref model).

In all the models explored, small mass subhaloes are always below the threshold for detection, and their presence could be revealed only through the enhancement of the diffuse foreground emission. The different predicted ratio between diffuse emission and number of detected haloes for the models we have considered, would provide precious information on the underlying cosmology, in case of positive detection. 

\section{Acknowledgments}
We are indebted to J. Bullock for help and discussion. 
We thank J. Diemand, S. Koushiappas and G. Tormen for useful discussions, comments and suggestions.

\end{document}